\documentclass[12pt,english]{article}
\usepackage[T1]{fontenc}
\usepackage[latin9]{inputenc}
\usepackage{amsmath}
\usepackage{amssymb}
\usepackage{graphicx}
\usepackage{setspace}
\makeatletter

\newcommand*\LyXThinSpace{\,\hspace{0pt}}
\providecommand{\tabularnewline}{\\}

\usepackage{braket}
 
\usepackage{fullpage}

\makeatother

\usepackage{babel}
\begin{document}

\thispagestyle{empty} \vskip 3em \begin{center} {{\bf \Large Coriolis terms in Skyrmion Quantization }}  \\[10mm]
{\bf \large J.~I. Rawlinson\footnote{email: jir25@damtp.cam.ac.uk}} \\[1pt] \vskip 1em {\it  Department of Applied Mathematics and Theoretical Physics,\\ University of Cambridge, \\ Wilberforce Road, Cambridge CB3 0WA, U.K.} \vspace{12mm}
\end{center}

\begin{abstract}
We consider the problem of quantizing a Skyrmion which is allowed
to vibrate, rotate and isorotate. Previous approaches have neglected
the interactions between vibrations and zero modes (analogous to so-called
\emph{Coriolis terms} in the molecular physics literature). A new
formalism incorporating these interactions is introduced, inspired
by a principal bundle approach to deformable-body dynamics. We quantize
the $B=4$ and \textbf{$B=7$} Skyrmions and compare the results to
observed nuclear properties of Helium-4 and the Lithium-7/Beryllium-7
isospin doublet.
\end{abstract}

\section{Introduction}

In the Skyrme model approach to nuclear physics, atomic nuclei appear
as topological solitons in a nonlinear field theory of pions. These
topological solitons are known as Skyrmions. Given a Skyrmion, one
may use insights from the field theory dynamics to identify a small
number of collective coordinates which are relevant at low energies.
The collective coordinates chosen will typically include both rotations
and isorotations of the Skyrmion together with other, shape-deforming,
degrees of freedom: the nucleus is viewed as a deformable body which
is free to rotate in space as well as isorotate in isospace. These
degrees of freedom of the Skyrmion are then quantized, hopefully giving
a reasonable description of the corresponding nucleus.

Naively one may hope to separate out the zero modes (rotations and
isorotations, corresponding to the action of the symmetry group $SU(2)_{\mathrm{spin}}\times SU(2)_{\mathrm{isospin}}$
on the Skyrmion) when quantizing this system. In some cases a complete
factorisation is possible, but generally one has to live with interactions
between zero modes and shape-deforming degrees of freedom. This kind
of interaction is already understood in the molecular physics literature,
where Coriolis effects are known to play an important role in rovibrational
spectra. Our situation is a slight generalisation: a molecule can
rotate and vibrate, but a Skyrmion can additionally isorotate. Mathematically
it is not difficult to incorporate isorotations, provided we have
a clear understanding of the usual Coriolis effects. The framework
introduced in \cite{littlejohn} (amongst others), which formulates
the problem in terms of principal bundles, is our preferred approach
and easily extends to include isorotations.

In section 2 we set up the general formalism before exploring various
applications in the following sections. In section 3 we consider small
vibrations of a Skyrmion and show how the problem simplifies in this
case. Using these ideas, we compute the quantum spectrum of a vibrating
and rotating \textbf{$B=4$ }Skyrmion (with cubic symmetry) in section
4, finding good agreement with the observed excited states of the
$\alpha$-particle. Finally we study the lowest-frequency vibration
of the $B=7$ Skyrmion, leading to a suggestion that the surprisingly
low energy of the Lithium-7/Beryllium-7 spin $\frac{3}{2}$ ground
state may be in part due to an isospin Coriolis effect.

\section{Quantization of Skyrmions}

\subsection*{Skyrme Lagrangian}

Pion fields $\mathbf{\pi}(x,t)$ are combined into an $SU(2)$-valued
field $U:\mathbb{R^{\mathrm{4}}\rightarrow}SU(2)$
\begin{equation}
U(x,t)=\sigma(x,t)\mathbb{I}_{2}+i\pi(x,t)\cdot\mathbb{\tau}\label{pionfields}
\end{equation}
and the Lagrangian defining the classical field theory is (in Skyrme
units)
\begin{equation}
L=\int d^{3}x\left[\frac{1}{2}\text{Tr}(L_{\mu}L^{\mu})+\frac{1}{16}\text{Tr}([L_{\mu},L_{\nu}][L^{\mu},L^{\nu}])+m^{2}Tr(U-\mathbb{I}_{2})\right]\label{eq:lagrangian}
\end{equation}
with $L_{\mu}=U^{\dagger}\partial_{\mu}U.$ Isospin symmetry corresponds
to transformations $U\rightarrow A^{\dagger}UA$ for any constant
matrix $A\in SU(2)$. Static soliton solutions are known as Skyrmions.
They are classified by a topological degree $B\in\mathbb{Z}$ which
is identified with the baryon number of the nucleus.

\subsection*{Restricted configuration space}

Given a Skyrmion, we are often interested in constructing a restricted
configuration space $\mathcal{C}$ of deformations. $\mathcal{C}$
should in principle capture the field configurations which are relevant
at low energies. A natural first choice is given by the rigid-body
approximation: only rotations and isorotations of the Skyrmion are
included. One then quantizes geodesic motion on the corresponding
submanifold $\mathcal{C}\simeq SU(2)_{\mathrm{}}\times SU(2)$ with
respect to the induced metric coming from the full field theory. The
resulting problem is equivalent to a (generalised) rigid rotor, with
quantum states classified by spin and isospin. Comparisons to nuclear
data have been promising in many cases, but recent work suggests that
to model real nuclei it is necessary to take additional deformations
of the Skyrmion into account: we need to include more than just the
zero modes. One can study vibrations of Skyrmions and find their normal
modes \cite{halcrow}. Then a natural next step beyond rigid-body
quantization is to include those modes with the lowest non-zero frequency
(the first $N$ of them, say). Within a harmonic approximation we
can think of the resulting configuration space as $\mathcal{C}\simeq SU(2)_{\mathrm{}}\times SU(2)\times\mathbb{R}^{N}$.
More generally we may be interested in larger collective motions (not
just small vibrations). Recent work on Carbon-12 \cite{rawlinson}
involved a configuration space of the form $\mathcal{C}\simeq SU(2)\times SU(2)\times\Gamma$
where $\Gamma$ has the structure of a graph, while in \cite{king}
Oxygen-16 was modelled by motion on $\mathcal{C}\simeq SU(2)\times SU(2)\times\mathcal{M}$
with $\mathcal{M}$ a quotient of a six-punctured sphere.

All of the examples given so far have the product structure $\mathcal{C}\simeq SU(2)\times SU(2)\times\mathcal{C}_{\mathrm{shapes}}$,
but one could imagine a restricted configuration space $\mathcal{C}$
which includes zero modes (generated by $SU(2)\times SU(2)$) but
is not globally a product. $\mathcal{C}$ should really be thought
of as a principal $SU(2)\times SU(2)$-bundle, with rotations and
isorotations generating the fibres. Locally it will be a product but
this might not be true globally.

\subsection*{Quantum Hamiltonian}

For clarity, we will at first ignore isorotations. Our configuration
space comes with an action of the rotational symmetry group $SU(2)$,
and we can think of the configuration space $\mathcal{C}$ as a principal
$SU(2)$-bundle $\pi:\mathcal{C}\rightarrow\mathcal{C}_{\mathrm{shapes}}$
with rotations generating the fibres. For every point in shape space
$\mathcal{C}_{\mathrm{shapes}}$ there is an open neighbourhood $V\subseteq\mathcal{C}_{\mathrm{shapes}}$
containing the point such that $\pi^{-1}\left(V\right)$ can be identified
with $SU(2)\times V$ (one should think of this as making a particular
choice of reference orientation for each fibre). Working locally,
we think of a point in configuration space as a pair $\left(\theta_{i},s_{j}\right)$
with $s_{j}$, the coordinates on $V\subseteq\mathcal{C}_{\mathrm{shapes}}$,
specifying the shape of the field configuration and with $\theta_{i}$
Euler angles parametrising its orientation in space. Our configuration
space inherits a metric $\tilde{g}$ from the full Skyrme field theory.
$SU(2)$ symmetry implies that this inherited metric must be symmetric
under (left) translations in the $SU(2)$ factor. Thus the most general
form of the inherited metric is
\begin{equation}
\tilde{g}=\begin{pmatrix}\mathbb{\sigma} & ds_{i}\end{pmatrix}\begin{pmatrix}\Lambda & \Lambda\mathbf{A}_{j}\\
\mathbf{A}_{i}^{T}\Lambda & g_{ij}+\mathbf{A}_{i}\cdot\Lambda\cdot\mathbf{A}_{j}
\end{pmatrix}\begin{pmatrix}\sigma\\
ds_{j}
\end{pmatrix}\label{eq:metricfull}
\end{equation}
where the $\sigma=\left(\sigma_{1},\sigma_{2},\sigma_{3}\right)$
are left-invariant one forms on $SU(2)$ and where $\Lambda$, $\mathbf{A}_{i}$
and $g_{ij}$ only depend on the shape coordinates $s_{i}.$ Note
that we have suppressed the index on $\sigma$, that $\Lambda$ is
a $3\times3$ matrix, and that a bold font is used to indicate that
$\mathbf{A}_{i}$ is a $3$-component vector for each $i$. The suggestive
notation $\mathbf{A}_{j}$ has been used as it will turn out that
this corresponds to a particular connection on the principal bundle
$\mathcal{C}$. We now construct a quantum Hamiltonian by computing
the Laplace-Beltrami operator on $\mathcal{C}$. Recall that the Laplace-Beltrami
operator $\Delta$ corresponding to a metric $G$ has an expression
in local coordinates
\begin{eqnarray}
\Delta f & = & \frac{1}{\sqrt{\left|G\right|}}\partial_{i}\left(\sqrt{\left|G\right|}G^{ij}\partial_{j}f\right).\label{laplacian}
\end{eqnarray}
 For the calculation of $\Delta$ it is useful to note that $\tilde{g}$
can be rewritten as
\begin{eqnarray}
\tilde{g} & = & \begin{pmatrix}d\theta & ds_{i}\end{pmatrix}G\begin{pmatrix}d\theta\\
ds_{j}
\end{pmatrix}\label{metrictilde}
\end{eqnarray}
(here we closely follow \cite{littlejohn}) where
\begin{equation}
G=\begin{pmatrix}\lambda^{T} & 0\\
\mathbf{A}_{i}^{T} & I
\end{pmatrix}\begin{pmatrix}\Lambda & 0\\
0 & g_{ij}
\end{pmatrix}\begin{pmatrix}\lambda & \mathbf{A}_{j}\\
0 & I
\end{pmatrix}.\label{metriccoord}
\end{equation}
$\lambda$ is the matrix which captures the relationship between the
left-invariant one forms $\sigma$ and the (coordinate) one forms
$d\theta$. Thus we can compute $\left|G\right|=\left|\lambda\right|^{2}\left|\Lambda\right|\left|g_{ij}\right|$
and then use the expression (\ref{laplacian}) to obtain a quantum
Hamiltonian 
\begin{equation}
\mathcal{H}=\frac{1}{2}\mathbf{L}\cdot\Lambda^{-1}\cdot\mathbf{L}+\frac{1}{2}\left(p_{i}-\mathbf{L}\cdot\mathbf{A}_{i}\right)g_{ij}^{-1}\left(p_{j}-\mathbf{L}\cdot\mathbf{A}_{j}\right)+V_{2}\left(s\right)+V(s)\label{eq:ham}
\end{equation}
where we have included both the kinetic term $-\hbar^{2}\Delta$ and
a potential $V(s)$ on configuration space. $\mathbf{L}$ is the (usual)
body-fixed angular momentum operator familiar from rigid-body theory
($\mathbf{J}$ will denote the space-fixed angular momentum operator)
and $p_{i}=-i\hbar\frac{\partial}{\partial s_{i}}$. Also appearing
in the kinetic term is
\begin{equation}
V_{2}\left(s\right)=\frac{\hbar^{2}}{2}\left(\left|\Lambda\right|\left|g_{ij}\right|\right)^{-\frac{1}{4}}\partial_{i}\left(g_{ij}^{-1}\partial_{j}\left(\left|\Lambda\right|\left|g_{ij}\right|\right)^{\frac{1}{4}}\right).\label{potent}
\end{equation}

\subsection*{Effective problem on $\mathcal{C}_{\mathrm{shapes}}$}

Exploiting rotational symmetry, we can classify the energy eigenstates
of (\ref{eq:ham}) by $J$ (where $J\left(J+1\right)$ is the eigenvalue
of $\mathbf{J}^{2}$ in the usual way) and $J_{3}$. Recall from rigid-body
theory that a complete set of commuting operators for the rotational
part of the problem is given by $\mathbf{J}^{2},J_{3},L_{3}$ and
so within a particular $\left(J,J_{3}\right)$ sector we can expand
the total wavefunction
\begin{equation}
\Psi=\sum_{L_{3}=-J}^{+J}\chi_{L_{3}}(s)\ket{JJ_{3}L_{3}}.\label{eq:}
\end{equation}
Within this sector, we see that $\Psi$ can be thought of as a complex
vector-valued function $\begin{pmatrix}\chi_{-J}(s)\\
\vdots\\
\chi_{J}(s)
\end{pmatrix}$ on $V\subseteq\mathcal{C}_{\mathrm{shapes}}$. Of course, we have
only been working locally, i.e. in some patch of $\mathcal{C}$ which
looks like a product $SU(2)\times V$. The total wavefunction, defined
on all of $\mathcal{C}$, isn't a vector-valued \emph{function} on
the base space but is more precisely a \emph{section} of a (complex)
vector bundle of rank $2J+1$. These two notions coincide for the
case of trivial bundles. In the more general case, we would work with
functions in separate patches and then impose appropriate conditions
on the overlaps to ensure they give a genuine section.

Substituting the expansion for $\Psi$ above into the Hamiltonian,
we obtain the Schrodinger equation
\begin{eqnarray}
\frac{1}{2}\mathbf{L}\cdot\Lambda^{-1}\cdot\mathbf{L}\begin{pmatrix}\chi_{-J}(s)\\
\vdots\\
\chi_{J}(s)
\end{pmatrix}+\frac{1}{2}\left(p_{i}-\mathbf{L}\cdot\mathbf{A}_{i}\right)g_{ij}^{-1}\left(p_{j}-\mathbf{L}\cdot\mathbf{A}_{j}\right)\begin{pmatrix}\chi_{-J}(s)\\
\vdots\\
\chi_{J}(s)
\end{pmatrix}\label{eq:-1}\\
+\left(V_{2}\left(s\right)+V(s)-E\right)\begin{pmatrix}\chi_{-J}(s)\\
\vdots\\
\chi_{J}(s)
\end{pmatrix} & = & 0\nonumber 
\end{eqnarray}
where now the operators $\mathbf{L}$ act by matrix multiplication.
This is the effective problem on $\mathcal{C}_{\mathrm{shapes}}$.
It is equivalent to the motion of a particle on $\mathcal{C}_{\mathrm{shapes}}$
coupled to an $SU(2)$ gauge field, with the particle transforming
in the $\left(2J+1\right)$-dimensional irrep of the gauge group and
with the gauge field (or connection) corresponding to $\mathbf{A}_{i}$.
Gauge transformations are equivalent to redefining our choice of reference
orientation for each $s\in\mathcal{C}_{\mathrm{shapes}}$. Note that
the rotational motion influences the motion on $\mathcal{C}_{\mathrm{shapes}}$
through the familiar minimal coupling $p_{j}-\mathbf{L}\cdot\mathbf{A}_{j}$
of the momentum $p_{j}$ to the gauge field. To completely separate
out rotational motion would require us to find a gauge where $\mathbf{A}_{i}$
vanishes. $\mathbf{A}_{i}$, while gauge dependent, has gauge-invariant
properties such as (possibly non-vanishing) curvature. The curvature
of $\mathbf{A}_{i}$ can therefore be viewed as an obstruction to
complete separation of rotational motion from the other degrees of
freedom.

\subsection*{Including isospin}

The above derivation is easily modified to include the possibility
of isospin. Once again the metric must take the form
\begin{equation}
\tilde{g}=\begin{pmatrix}\mathbb{\sigma} & ds_{i}\end{pmatrix}\begin{pmatrix}\Lambda & \Lambda\mathbf{A}_{j}\\
\mathbf{A}_{i}^{T}\Lambda & g_{ij}+\mathbf{A}_{i}\cdot\Lambda\cdot\mathbf{A}_{j}
\end{pmatrix}\begin{pmatrix}\sigma\\
ds_{j}
\end{pmatrix}\label{eq:-2}
\end{equation}
where now $\sigma=\left(\sigma_{1}^{J},\sigma_{2}^{J},\sigma_{3}^{J},\sigma_{1}^{I},\sigma_{2}^{I},\sigma_{3}^{I}\right)$
includes both left-invariant one forms $\left(\sigma_{1}^{J},\sigma_{2}^{J},\sigma_{3}^{J}\right)$
associated with rotations and $\left(\sigma_{1}^{I},\sigma_{2}^{I},\sigma_{3}^{I}\right)$
associated with isorotations. $\Lambda$ and $\mathbf{A}_{i}$ have
now become a $6\times6$ matrix and (for each $i$) a $6$-component
vector respectively. One ends up with the Hamiltonian
\begin{equation}
\mathcal{H}=\frac{1}{2}\begin{pmatrix}\mathbf{L}\\
\mathbf{K}
\end{pmatrix}\cdot\Lambda^{-1}\cdot\begin{pmatrix}\mathbf{L}\\
\mathbf{K}
\end{pmatrix}+\frac{1}{2}\left(p_{i}-\begin{pmatrix}\mathbf{L}\\
\mathbf{K}
\end{pmatrix}\cdot\mathbf{A}_{i}\right)g_{ij}^{-1}\left(p_{j}-\begin{pmatrix}\mathbf{L}\\
\mathbf{K}
\end{pmatrix}\cdot\mathbf{A}_{j}\right)+V_{2}\left(s\right)+V(s)\label{eq:-3}
\end{equation}
where 
\begin{equation}
V_{2}=\frac{\hbar^{2}}{2}\left(\left|\Lambda\right|\left|g_{ij}\right|\right)^{-\frac{1}{4}}\partial_{i}\left(g_{ij}^{-1}\partial_{j}\left(\left|\Lambda\right|\left|g_{ij}\right|\right)^{\frac{1}{4}}\right).\label{eq:-4}
\end{equation}
We will make use of this Hamiltonian later, but for now we will go
back to only including rotations.

\section{Equilateral triangle in $\mathbb{R}^{3}$}

We will be interested in small vibrations of Skyrmions, ultimately
applying the above ideas to quantization of the $B=4$ and $B=7$
Skyrmions. But let us start with a simpler problem which illustrates
the main ideas: as a model for a Skyrmion, consider an equilateral
triangular arrangement of point particles in $\mathbb{R}^{3}$, with
particle $i$ having mass $m$ and position vector $\mathbf{r}_{i}$.
We imagine these are attached by identical springs. This (equilateral)
arrangement has symmetry group $D_{3h}$, and so its vibrations can
be classified by irreducible representations (irreps) of this group.
There are three normal modes (not including zero modes), which split
into the irreps $A\oplus E'$ under the action of $D_{3h}$. For the
spring model, the $E^{'}$ vibration has lowest frequency with $\frac{\omega_{E^{'}}}{\omega_{A}}=\frac{1}{\sqrt{2}}$.
Suppose we are interested in a configuration space $\mathcal{C}\simeq SO(3)\times\mathbb{R}^{2}$
which includes only this doubly-degenerate vibration ($E'$) together
with rotations. This is clearly a trivial bundle. We will use coordinates
$\mathbf{s}=\left(s_{1},s_{2}\right)$ on $C_{\mathrm{shapes}}$,
and Euler angles $\theta^{i}$ to specify orientation. Let $d$ be
the distance from each particle to the centre of mass in the equilibrium
configuration, and work in units where $\hbar=1,m=1,d=1.$ Let the
coordinates $\left(\theta^{i}=0,\mathbf{s}\right)$ correspond to
the configuration
\begin{equation}
\begin{array}{c}
\mathbf{r}_{1}=\begin{pmatrix}0\\
1\\
0
\end{pmatrix}+s_{1}\begin{pmatrix}0\\
\frac{1}{\sqrt{3}}\\
0
\end{pmatrix}+s_{2}\begin{pmatrix}\frac{1}{\sqrt{3}}\\
0\\
0
\end{pmatrix}\\
\\
\mathbf{r}_{2}=\begin{pmatrix}\frac{\sqrt{3}}{2}\\
-\frac{1}{2}\\
0
\end{pmatrix}+s_{1}\begin{pmatrix}-\frac{1}{2}\\
-\frac{1}{2\sqrt{3}}\\
0
\end{pmatrix}+s_{2}\begin{pmatrix}-\frac{1}{2\sqrt{3}}\\
\frac{1}{2}\\
0
\end{pmatrix}\\
\\
\mathbf{r}_{3}=\begin{pmatrix}-\frac{\sqrt{3}}{2}\\
-\frac{1}{2}\\
0
\end{pmatrix}+s_{1}\begin{pmatrix}\frac{1}{2}\\
-\frac{1}{2\sqrt{3}}\\
0
\end{pmatrix}+s_{2}\begin{pmatrix}-\frac{1}{2\sqrt{3}}\\
-\frac{1}{2}\\
0
\end{pmatrix}.
\end{array}\label{eq:-5}
\end{equation}
This is our gauge choice. A general configuration $\left(\theta^{i},\mathbf{s}\right)$
with $\theta^{i}\neq0$ can be deduced from a rotation of the corresponding
reference configuration $\left(0,\mathbf{s}\right)$. We will assume
$V(\mathbf{s})=\frac{1}{2}\omega^{2}\mathbf{s}^{2}$. We can compute
the metric induced from the Euclidean metric on $\mathbb{R}^{9}$
(three point particles) which leads, by comparison to the expression
\begin{equation}
\tilde{g}=\begin{pmatrix}\mathbb{\sigma} & ds_{i}\end{pmatrix}\begin{pmatrix}\Lambda & \Lambda\mathbf{A}_{j}\\
\mathbf{A}_{i}^{T}\Lambda & g_{ij}+\mathbf{A}_{i}\cdot\Lambda\cdot\mathbf{A}_{j}
\end{pmatrix}\begin{pmatrix}\sigma\\
ds_{j}
\end{pmatrix},\label{eq:-6}
\end{equation}
 to 
\begin{equation}
\Lambda=\begin{pmatrix}\frac{3}{2}+\sqrt{3}s_{1}+\frac{1}{2}\left(s_{1}^{2}+s_{2}^{2}\right) & -\sqrt{3}s_{2} & 0\\
-\sqrt{3}s_{2} & \frac{3}{2}-\sqrt{3}s_{1}+\frac{1}{2}\left(s_{1}^{2}+s_{2}^{2}\right) & 0\\
0 & 0 & 3+s_{1}^{2}+s_{2}^{2}
\end{pmatrix},\label{eq:-7}
\end{equation}
\begin{eqnarray}
g_{ij} & = & \frac{1}{3+s_{1}^{2}+s_{2}^{2}}\begin{pmatrix}3+s_{1}^{2} & s_{1}s_{2}\\
s_{1}s_{2} & 3+s_{2}^{2}
\end{pmatrix},\label{eq:-8}
\end{eqnarray}
and
\begin{equation}
\begin{array}{ccc}
\mathbf{A}_{1}=\begin{pmatrix}0\\
0\\
\frac{s_{2}}{3+s_{1}^{2}+s_{2}^{2}}
\end{pmatrix} &  & \mathbf{A}_{2}=\begin{pmatrix}0\\
0\\
-\frac{s_{1}}{3+s_{1}^{2}+s_{2}^{2}}
\end{pmatrix}\end{array}.\label{eq:-9}
\end{equation}
We can already see that the gauge field takes a familiar form for
small $\left(s_{1},s_{2}\right)$: we have $\mathbf{A}_{1}\sim\begin{pmatrix}0\\
0\\
\frac{s_{2}}{3}
\end{pmatrix}$, $\mathbf{A}_{2}\sim\begin{pmatrix}0\\
0\\
-\frac{s_{1}}{3}
\end{pmatrix}$ and so the effective motion on $\mathcal{C}_{\mathrm{shapes}}$ will
appear as if coupled to a constant magnetic field (of strength $\frac{1}{3}L_{3}$)
pointing out of the $\left(s_{1},s_{2}\right)$-plane. For this example
it is also simple to compute 
\begin{eqnarray}
V_{2}\left(s\right) & = & \frac{1}{2}\left(\left|\Lambda\right|\left|g_{ij}\right|\right)^{-\frac{1}{4}}\partial_{i}\left(g_{ij}^{-1}\partial_{j}\left(\left|\Lambda\right|\left|g_{ij}\right|\right)^{\frac{1}{4}}\right)\label{eq:-10}\\
 & = & \frac{1}{2}\frac{-6+s_{1}^{2}+s_{2}^{2}}{\left(3-s_{1}^{2}-s_{2}^{2}\right)^{2}}.\nonumber 
\end{eqnarray}
Recall that the full quantum Hamiltonian is
\begin{equation}
\mathcal{H}=\frac{1}{2}\mathbf{L}\cdot\Lambda^{-1}\cdot\mathbf{L}+\frac{1}{2}\left(p_{i}-\mathbf{L}\cdot\mathbf{A}_{i}\right)g_{ij}^{-1}\left(p_{j}-\mathbf{L}\cdot\mathbf{A}_{j}\right)+V_{2}\left(s\right)+V(s).\label{eq:-11}
\end{equation}
We now make the following approximation: assume that the vibrational
frequency $\omega$ is large so that the most important terms in the
above Hamiltonian give a harmonic oscillator
\begin{equation}
\mathcal{H}_{0}=\frac{1}{2}\left(p_{1}^{2}+p_{2}^{2}\right)+\frac{1}{2}\omega^{2}\left(s_{1}^{2}+s_{2}^{2}\right).\label{eq:-12}
\end{equation}
We will expand the full Hamiltonian in $\frac{1}{\omega}$, keeping
the leading corrections to the $\mathcal{H}_{0}$ system. Note that
$\mathcal{H}_{0}$ has eigenvalues $\sim\omega$. Also note that,
schematically, $s^{2}\sim\frac{1}{\omega}$ and $p^{2}\sim\omega$
for the harmonic oscillator from which the orders of other terms in
$\mathcal{H}$ can be deduced. Expanding out the full Hamiltonian,
we have
\begin{equation}
\begin{array}{ccccccc}
\mathcal{H} & = & \mathcal{H}_{0} & + & \underbrace{\frac{1}{3}\left(\mathbf{L}^{2}-\frac{1}{2}L_{3}^{2}\right)+\frac{1}{3}L_{3}J_{s}+\frac{1}{6}J_{s}^{2}-\frac{1}{3}} & + & \ldots\\
 &  & \sim\omega^{1} &  & \sim\omega^{0} &  & \sim\mathrm{higher}
\end{array}\label{eq:-13}
\end{equation}
where $J_{s}=s_{1}p_{2}-s_{2}p_{1}$ is an operator which will be
referred to as the vibrational angular momentum. To this order, the
only effect of $V_{2}\left(s\right)$ is to contribute an additive
constant (here, $-\frac{1}{3}$) to the Hamiltonian. This will be
the case more generally and so we will neglect $V_{2}\left(s\right)$
in later examples. So the terms that remain are $\mathcal{H}_{0}$
(a harmonic oscillator corresponding to vibrations), $\frac{1}{3}\left(\mathbf{L}^{2}-\frac{1}{2}L_{3}^{2}\right)$
(the familiar rigid-body Hamiltonian, corresponding to rotations)
and finally the term $\frac{1}{3}L_{3}J_{s}+\frac{1}{6}J_{s}^{2}$
which comes from the gauge field. It gives the leading correction
due to rotation-vibration coupling. This is referred to as a \emph{Coriolis
term }in the molecular physics literature.

\subsection*{Symmetry arguments}

Before moving on, let's reflect on what we have done in this example.
The coordinates $\left(\theta^{i},\mathbf{s}\right)$ were actually
carefully chosen so that the metric took the form
\begin{equation}
\tilde{g}=\begin{pmatrix}\mathbb{\sigma} & ds_{i}\end{pmatrix}\begin{pmatrix}\Lambda_{0} & \Lambda_{0}\mathbf{A}_{j}\\
\mathbf{A}_{i}^{T}\Lambda_{0} & \delta_{ij}
\end{pmatrix}\begin{pmatrix}\sigma\\
ds_{j}
\end{pmatrix}\label{eq:-14}
\end{equation}
with, to the order we are interested in, $\Lambda=\Lambda_{0}$ a
constant matrix (the moment of inertia tensor for the equilibrium
configuration) and bottom-right entry $\delta_{ij}$ (normal coordinates
for the vibration) and with the off-diagonal entry $\Lambda\mathbf{A}_{i}$
vanishing at the equilibrium configuration (this says that rotations
and vibrations are orthogonal at the equilibrium configuration). Then,
to the order we are interested in, $\mathbf{A}_{i}$ is linear in
the shape coordinates. Now recall that the equilibrium configuration
has $D_{3h}$ symmetry and that the vibration we are interested in
transforms in the $E'$ representation of $D_{3h}$, $\rho_{\mathrm{vib}}\cong E'$.
In particular, the metric $\tilde{g}$ enjoys a $D_{3h}$ symmetry
and so the gauge field $\mathbf{A}_{i}$ is not just an \emph{arbitrary}
linear function of the shape coordinates but corresponds to a singlet
of $D_{3h}$ under an action of $D_{3h}$ isomorphic to $\rho_{\mathrm{vib}}\otimes\rho_{\mathrm{vib}}\otimes\rho_{\mathrm{rot}}$
where $\rho_{\mathrm{rot}}$ denotes the representation in which rotations
$\left(R_{x},R_{y},R_{z}\right)$ transform under $D_{3h}$ (for our
example $\rho_{\mathrm{rot}}=A_{2}'\oplus E''$). This observation
is equivalent to \emph{Jahn's rule}, which is known in molecular physics
as a necessary condition for the existence of non-trivial first-order
Coriolis terms \cite{jahn}. In the present case, a simple character
theory calculation shows that $\rho_{\mathrm{vib}}\otimes\rho_{\mathrm{vib}}\otimes\rho_{\mathrm{rot}}$
contains precisely one copy of the trivial irrep of $D_{3h}$. So
the gauge field $\mathbf{A}_{i}$ is determined by a single constant
$\eta$. It has to transform trivially under $\rho_{\mathrm{vib}}\otimes\rho_{\mathrm{vib}}\otimes\rho_{\mathrm{rot}}$,
which in this case means that the $\mathbf{A}_{i}$ must satisfy
\begin{equation}
\forall g\in D_{3h}:\rho_{\text{rot}}\left(g\right)\left(\rho_{\text{vib}}\left(g\right)\right)_{ik}\mathbf{A}_{k}\left(\left(\rho_{\text{vib}}\left(g\right)\right)_{jl}^{-1}s_{l}\right)=\mathbf{A}_{i}\left(s_{j}\right)\label{eq:-32}
\end{equation}
so that

\begin{equation}
\begin{array}{ccc}
\mathbf{A}_{1}=\eta\begin{pmatrix}0\\
0\\
s_{2}
\end{pmatrix} &  & \mathbf{A}_{2}=\eta\begin{pmatrix}0\\
0\\
-s_{1}
\end{pmatrix}.\end{array}\label{eq:-15}
\end{equation}
The only reason to do the explicit calculation of the previous section
was to determine that $\eta=\frac{1}{3}$. We might more generally
take $\eta$ to be a free parameter. This insight will prove useful
in situations where it is not so easy to compute the gauge field explicitly,
and all we have is knowledge of the relevant symmetry group together
with the transformation properties of the vibration.

\section{$B=4$ Skyrmion and the $\alpha$-particle}

\begin{figure}[h]
\centering{}\includegraphics[scale=0.1]{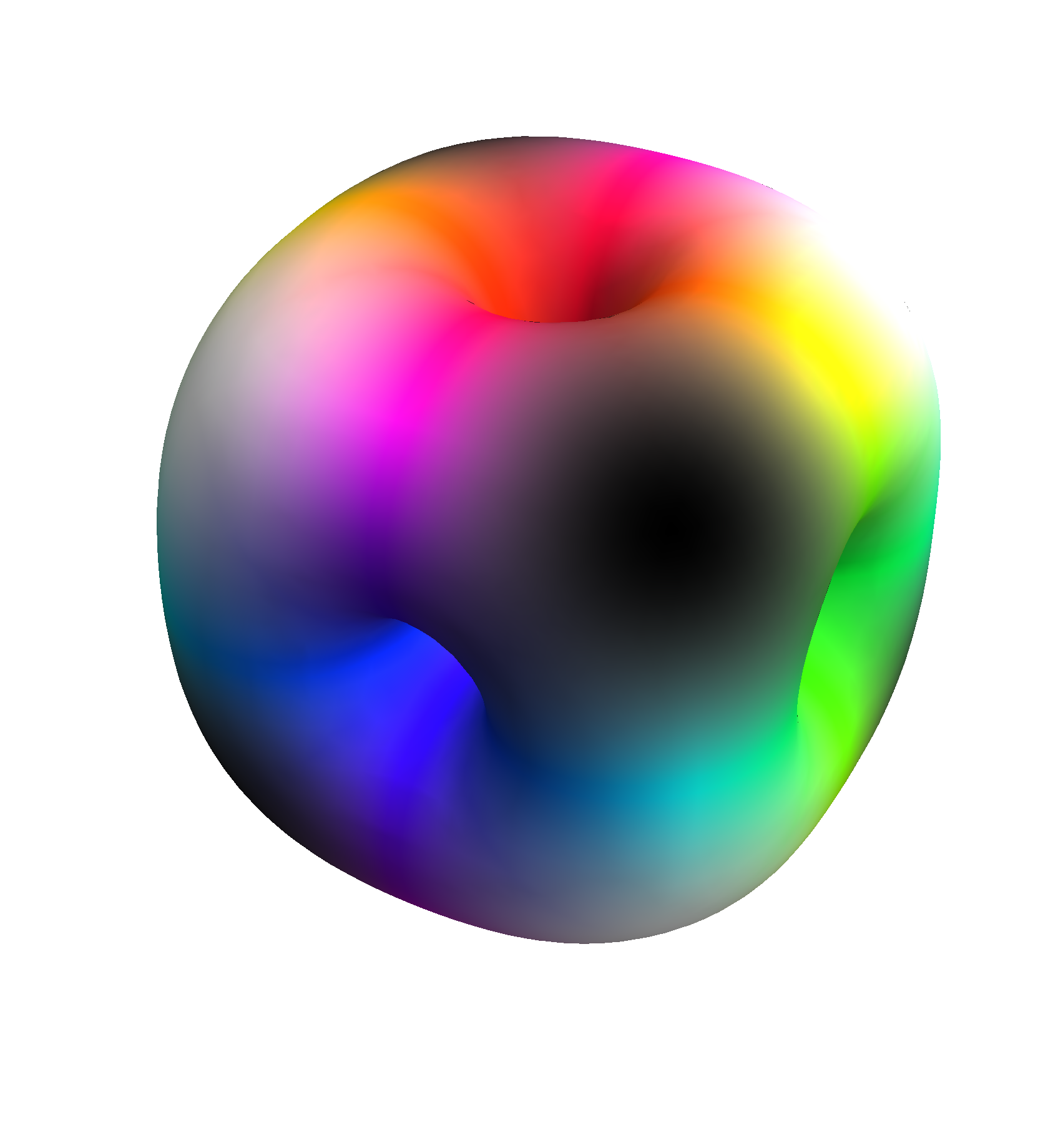}\caption{B=4 Skyrmion with $O_{h}$ symmetry. Figure courtesy of Dankrad Feist.}
\label{b4}
\end{figure}
We now apply our insights from the previous section to the problem
of a vibrating and rotating Skyrmion. The minimal energy $B=4$ Skyrmion
has $O_{h}$ symmetry and is illustrated in Figure \ref{b4}. The
isospin $0$ quantum states of this Skyrmion correspond to the $\alpha$-particle.
In \cite{manko} the authors performed rigid-body quantization of
the $O_{h}$-symmetric $B=4$ Skyrmion, finding a ground state with
spin $J=0$ and a first excited state with $J=4$. The spin $4$ excitation
(at roughly $40$ MeV) has not yet been experimentally observed, however
there are numerous observed excited states with lower spin in the
$20-30$ MeV range \cite{tunl} which are not captured by the rigid-body
picture. The $O_{h}$ symmetry group of the rigid $B=4$ Skyrmion
is too large to allow such excitations (which have spins $0$, $1$
and $2$) and so the data suggests that vibrations must be included
if we want to describe these states.

\subsubsection*{Vibrations of the $B=4$ Skyrmion}

The lowest four vibrational modes \cite{halcrow,baskerville} of the
$B=4$ $O_{h}$-symmetric Skyrmion are listed in Table \ref{vibs}.
The associated frequences are those calculated in \cite{halcrow}
for a dimensionless pion mass of $m=1$. These vibrations have been
classified using $O_{h}$ representation theory. As a group, $O_{h}$
is generated by a $3$-fold rotation $C_{3}$, 4-fold rotation $C_{4}$
together with an inversion element $-I$. $O_{h}$ has $10$ irreps
with the corresponding character table given in Table \ref{ohchar}.
\begin{table}[h]
\begin{tabular}{c|c|c}
Frequency & Irrep of $O_{h}$ & Description\tabularnewline
\hline 
$0.46$ & $E^{+}$ & Two opposite faces pull away from each other\tabularnewline
 &  & to form two $B=2$ tori. In the other direction, four\tabularnewline
 &  & edges pull away to become four $B=1$ Skyrmions.\tabularnewline
\hline 
$0.48$ & $F_{2}^{+}$ & An opposing pair of square-symmetric\tabularnewline
 &  & faces deform to become rhombus-shaped.\tabularnewline
\hline 
$0.52$ & $A_{2}^{-}$ & Four vertices of the cube pull away, retaining\tabularnewline
 &  & tetrahedral symmetry. These then come in again and the\tabularnewline
 &  & other four vertices pull away to form the dual tetrahedron.\tabularnewline
\hline 
$0.62$ & $F_{2}^{-}$ & Two opposite edges from the same face\tabularnewline
 &  & pull away from the origin. On the opposite \tabularnewline
 &  & face, the perpendicular edges also pull away.\tabularnewline
\hline 
\end{tabular}

\caption{Vibrations of $B=4$ Skyrmion. Frequencies and descriptions from \cite{halcrow}.}
\label{vibs}
\end{table}
Following a similar approximation scheme to the previous section,
our aim is to compute the quantum spectrum of a vibrating and rotating
\textbf{$B=4$ }Skyrmion. We assume that the vibrations in Table \ref{vibs}
are the most important and neglect any other degrees of freedom. We
will ignore isorotations as we are interested in isospin 0 states
corresponding to the $\alpha$-particle. (We will include isorotations
when we look at the $B=7$ Skyrmion in the next section).
\begin{table}[h]
\begin{centering}
\begin{tabular}{c|c|c|c|c|c|c|c|c|c|c}
$O_{h}$ & $E$ & $8C_{3}$ & $6C_{2}$ & $6C_{4}$ & $3C_{2}=\left(C_{4}\right)^{2}$ & $i$ & $6S_{4}$ & $8S_{6}$ & $3\sigma_{h}$ & $6\sigma_{d}$\tabularnewline
\hline 
$A_{1}^{+}$ & $1$ & $1$ & $1$ & $1$ & $1$ & $1$ & $1$ & $1$ & $1$ & $1$\tabularnewline
\hline 
$A_{2}^{+}$ & $1$ & $1$ & $-1$ & $-1$ & $1$ & $1$ & $-1$ & $1$ & $1$ & $-1$\tabularnewline
\hline 
$E^{+}$ & $2$ & $-1$ & $0$ & $0$ & $2$ & $2$ & $0$ & $-1$ & $2$ & $0$\tabularnewline
\hline 
$F_{1}^{+}$ & $3$ & $0$ & $-1$ & $1$ & $-1$ & $3$ & $1$ & $0$ & $-1$ & $-1$\tabularnewline
\hline 
$F_{2}^{+}$ & $3$ & $0$ & $1$ & $-1$ & $-1$ & $3$ & $-1$ & $0$ & $-1$ & $1$\tabularnewline
\hline 
$A_{1}^{-}$ & $1$ & $1$ & $1$ & $1$ & $1$ & $-1$ & $-1$ & $-1$ & $-1$ & $-1$\tabularnewline
\hline 
$A_{2}^{-}$ & $1$ & $1$ & $-1$ & $-1$ & $1$ & $-1$ & $1$ & $-1$ & $-1$ & $1$\tabularnewline
\hline 
$E^{-}$ & $2$ & $-1$ & $0$ & $0$ & $2$ & $-2$ & $0$ & $1$ & $-2$ & $0$\tabularnewline
\hline 
$F_{1}^{-}$ & $3$ & $0$ & $-1$ & $1$ & $-1$ & $-3$ & $-1$ & $0$ & $1$ & $1$\tabularnewline
\hline 
$F_{2}^{-}$ & $3$ & $0$ & $1$ & $-1$ & $-1$ & $-3$ & $1$ & $0$ & $1$ & $-1$\tabularnewline
\hline 
\end{tabular}
\par\end{centering}
\caption{$O_{h}$ character table \cite{character}.}
\label{ohchar}
\end{table}

\subsubsection*{The $F_{2}^{-}$ vibration}

To start off, we consider just the triply degenerate $F_{2}^{-}$
vibration of the $B=4$ Skyrmion along with rotations, $\mathcal{C}\simeq SU(2)\times\mathbb{R}^{3}.$
As in the case of the equilateral triangle, we assume the vibrations
are small. The equilibrium configuration has symmetry group $O_{h}$,
which acts on physical space as follows:
\[
C_{4}:\left(x,y,z\right)\rightarrow\left(-y,x,z\right)
\]
\begin{equation}
C_{3}:\left(x,y,z\right)\rightarrow\left(y,z,x\right)\label{eq:-16}
\end{equation}
\[
-I:\left(x,y,z\right)\rightarrow\left(-x,-y,-z\right).
\]
Note that this action of $O_{h}$ is isomorphic to $F_{1}^{-}$. Introduce
coordinates $\left(\theta^{i},\mathbf{s}\right)$ such that the total
metric takes the form (to the order we are interested in)
\begin{equation}
\tilde{g}=\begin{pmatrix}\mathbb{\sigma} & ds_{i}\end{pmatrix}\begin{pmatrix}\Lambda_{0} & \Lambda_{0}\mathbf{A}_{i}\\
\mathbf{A}_{i}^{T}\Lambda_{0} & \delta_{ij}
\end{pmatrix}\begin{pmatrix}\sigma\\
ds_{j}
\end{pmatrix}\label{eq:-17}
\end{equation}
with 
\begin{equation}
\Lambda_{0}=\begin{pmatrix}\mathcal{I} & 0 & 0\\
0 & \mathcal{I} & 0\\
0 & 0 & \mathcal{I}
\end{pmatrix},\label{eq:-18}
\end{equation}
$\mathbf{A}_{i}$ linear and vanishing at the equilibrium configuration
$s_{1}=s_{2}=s_{3}=0.$ We still have some freedom in which vibrational
coordinates $\mathbf{s}=\left(s_{1},s_{2},s_{3}\right)$ we choose,
and we will choose them so that they transform under $O_{h}$ as follows:
\[
C_{4}:\left(s_{1},s_{2},s_{3}\right)\rightarrow\left(s_{2},-s_{1},-s_{3}\right)
\]
\begin{equation}
C_{3}:\left(s_{1},s_{2},s_{3}\right)\rightarrow\left(s_{2},s_{3},s_{1}\right)\label{eq:-19}
\end{equation}
\[
-I:\left(s_{1},s_{2},s_{3}\right)\rightarrow\left(-s_{1},-s_{2},-s_{3}\right).
\]
Note at this point that, unlike in the point particle example of the
previous section, we do not have explicit expressions for the Skyrme
field configurations corresponding to each $\mathbf{s}=\left(s_{1},s_{2},s_{3}\right)$.
However, it is always possible to pick coordinates so that the action
of $O_{h}$ is realised exactly as above (since the representation
$\rho_{\mathrm{vib}}$ of $O_{h}$ given in (\ref{eq:-19}) is indeed
isomorphic to $F_{2}^{-}$). As in the point particle model of the
previous section, it turns out that $\rho_{\mathrm{vib}}\otimes\rho_{\mathrm{vib}}\otimes\rho_{\mathrm{rot}}=F_{2}^{-}\otimes F_{2}^{-}\otimes F_{1}^{+}$
contains precisely one singlet, and that the gauge field is therefore
determined up to a single multiplicative scalar $\eta_{-}$: 
\begin{equation}
\begin{array}{ccccc}
\mathbf{A}_{1}=\frac{\eta_{-}}{2\mathcal{I}}\begin{pmatrix}0\\
-s_{3}\\
s_{2}
\end{pmatrix} &  & \mathbf{A}_{2}=\frac{\eta_{-}}{2\mathcal{I}}\begin{pmatrix}s_{3}\\
0\\
-s_{1}
\end{pmatrix} &  & \mathbf{A}_{3}=\frac{\eta_{-}}{2\mathcal{I}}\begin{pmatrix}-s_{2}\\
s_{1}\\
0
\end{pmatrix}\end{array}.\label{eq:-20}
\end{equation}
Substituting this into the general expression in (\ref{eq:-11}),
we arrive at the Hamiltonian
\begin{equation}
\mathcal{H\approx}\frac{1}{2}\mathbf{p}^{2}+\frac{1}{2}\omega_{F_{2}^{-}}^{2}\mathbf{s}^{2}+\frac{1}{2\mathcal{I}}\mathbf{L}^{2}-\frac{\eta_{-}}{2\mathcal{I}}\mathbf{L}\cdot\mathbf{J}_{\mathbf{s}}+\frac{\eta_{-}^{2}}{8\mathcal{I}}\mathbf{J}_{\mathbf{s}}^{2}\label{eq:-21}
\end{equation}
where $p_{i}=-i\hbar\frac{\partial}{\partial s_{i}}$ and $\mathbf{J}_{\mathbf{s}}=\mathbf{s}\times\mathbf{p}_{\mathbf{}}$.
A similar picture to that in equation (\ref{eq:-13}) emerges: we
have a harmonic oscillator system and a rigid-body system which are
coupled through the additional term $-\frac{\eta_{-}}{2\mathcal{I}}\mathbf{L}\cdot\mathbf{J}_{\mathbf{s}}+\frac{\eta_{-}^{2}}{8\mathcal{I}}\mathbf{J}_{\mathbf{s}}^{2}$
involving the usual body-fixed angular momentum $\mathbf{L}$ and
a vibrational angular momentum $\mathbf{J}_{\mathbf{s}}$. In principle
$\eta_{-}$ could be calculated from the Skyrme model given explicit
Skyrme field configurations, much like how we calculated $\eta=\frac{1}{3}$
in the preceeding (point particle) example. We will take it to be
a free parameter.

\subsubsection*{Computing the spectrum}

We are interested in the Hamiltonian
\begin{equation}
\mathcal{H=}\frac{1}{2}\mathbf{p}^{2}+\frac{1}{2}\omega_{F_{2}^{-}}^{2}\mathbf{s}^{2}+\frac{1}{2I}\mathbf{L}^{2}-\frac{\eta_{-}}{2I}\mathbf{L}\cdot\mathbf{J}_{\mathbf{s}}+\frac{\eta_{-}^{2}}{8I}\mathbf{J}_{\mathbf{s}}^{2}.\label{eq:-22}
\end{equation}
It will help to rewrite the Hamiltonian using the fact that, as $\mathbf{J}_{\mathbf{s}}$
and $\mathbf{L}$ commute, 
\begin{equation}
\mathbf{L}\cdot\mathbf{J}_{\mathbf{s}}=\frac{1}{2}\mathbf{L}^{2}+\frac{1}{2}\mathbf{J}_{\mathbf{s}}^{2}-\frac{1}{2}\mathbf{M}^{2}\label{eq:-40}
\end{equation}
where we have introduced a new angular momentum operator $\mathbf{M}=\mathbf{J}_{\mathbf{s}}-\mathbf{L}$.
(Note that $-\mathbf{L}$, not $+\mathbf{L}$, obeys the usual angular
momentum commutation relations: $\mathbf{L}$ is the vector of \emph{body-fixed}
angular momentum operators so its commutation relations differ by
a minus sign compared to \emph{space-fixed} angular momentum operators).
Then
\begin{equation}
\mathcal{H=}\frac{1}{2}\mathbf{p}^{2}+\frac{1}{2}\omega_{F_{2}^{-}}^{2}\mathbf{s}^{2}+\left(\frac{1}{2\mathcal{I}}-\frac{\eta_{-}}{4\mathcal{I}}\right)\mathbf{L}^{2}+\frac{\eta_{-}}{4\mathcal{I}}\mathbf{M}^{2}+\left(\frac{\eta_{-}^{2}}{8\mathcal{I}}-\frac{\eta_{-}}{4\mathcal{I}}\right)\mathbf{J}_{\mathbf{s}}^{2}.\label{eq:-23}
\end{equation}
Energy eigenstates $\Psi$ can be classified by $\mathbf{M}^{2},\mathbf{J}_{\mathbf{s}}^{2},\mathbf{L}^{2}$
and the vibrational phonon-number $N_{F_{2}^{-}}$, and additionally
by their transformation under the $O_{h}$ symmetry group, where $O_{h}$
acts on a state by transforming the vibrational coordinates, and then
performing a compensating rotation:
\begin{equation}
\Psi\rightarrow\rho_{\mathrm{rot}}\left(g\right)\otimes\rho_{\mathrm{vib}}\left(g\right)\Psi.\label{eq:-24}
\end{equation}
Explicitly, this action is generated by
\[
C_{4}:\Psi\rightarrow Pe^{-\frac{2\pi i}{4}\mathbf{n}_{4}\cdot\mathbf{M}}\Psi
\]
\begin{equation}
C_{3}:\Psi\rightarrow e^{-\frac{2\pi i}{3}\mathbf{n}_{3}\cdot\mathbf{M}}\Psi\label{eq:-25}
\end{equation}
\[
-I:\Psi\rightarrow P\Psi
\]
where $\mathbf{n}_{4}=\begin{pmatrix}0\\
0\\
1
\end{pmatrix}$, $\mathbf{n}_{3}=\frac{1}{\sqrt{3}}\begin{pmatrix}-1\\
-1\\
-1
\end{pmatrix}$ and where $P$ is the parity operation on the vibrational coordinates
$\mathbf{s}\rightarrow-\mathbf{s}.$ The Finkelstein-Rubinstein (F-R)
constraints tell us that physical states should be taken to transform
trivially under the action of $\left\langle C_{4},C_{3}\right\rangle \cong O\leq O_{h}$,
the subgroup consisting of rotations. Recall $O_{h}\cong O\times\mathbb{Z}_{2}$,
a direct product of groups, with $\mathbb{Z}_{2}$ the subgroup generated
by the parity operation $-I\in O_{h}$. So such representations fall
into two classes, $A_{1}^{+}$ or $A_{1}^{-}$, depending on their
transformation under the $\mathbb{Z}_{2}$. This determines the parity
of the state as $+$ or $-$. Within each fixed $\mathbf{M}^{2},\mathbf{J}_{\mathbf{s}}^{2},\mathbf{L}^{2},N_{F_{2}^{-}}$
sector, we compute the character of the action of $O_{h}$ and then
look for representations of type $A_{1}$. For example: suppose we
are interested in one-phonon states (i.e. states with one quantum
of vibrational energy). Such states have $J_{\mathbf{s}}=1.$ We might
look for states with $J=L=2$. Adding these angular momenta, we have
several possibilities for the total angular momentum $M=J_{\mathbf{s}}-L=3,2,1.$
So, if we are interested say in $M=3$, we have narrowed down to a
$7$-dimensional subspace. We now look at how this $7$-dimensional
subspace transforms under the action above, computing the associated
character $\chi$. We then find that $\left\langle \chi,A_{1}^{+}\right\rangle =0$
and $\left\langle \chi,A_{1}^{-}\right\rangle =1$ giving a single
negative parity $2^{-}$ state.

\subsubsection*{Other vibrations}

We now include the vibrations transforming as $F_{2}^{-},F_{2}^{+},A_{2}^{-}$,
treating the vibrational frequencies as free parameters, and fit the
resulting spectrum to data in the $<30$ MeV range. We could also
include the $E^{+}$ vibration but it turns out that including it
gives no improvement to the fit to experimental data. In fact it will
turn out that almost all of the data can be explained solely in terms
of $F_{2}^{-}$ and $F_{2}^{+}$ modes, with a higher frequency $A_{2}^{-}$
mode important for a couple of higher energy ($\sim28$ MeV) states.
The $F_{2}^{-}$ and $F_{2}^{+}$ can have Coriolis terms whereas
symmetry considerations exclude any non-trivial Coriolis term for
the $A_{2}^{-}$. This leads to the Hamiltonian
\begin{eqnarray}
\mathcal{H} & = & \frac{1}{2}\mathbf{p}_{\mathbf{s}}^{2}+\frac{1}{2}\omega_{F_{2}^{-}}^{2}\mathbf{s}^{2}+\frac{1}{2}\mathbf{p}_{\mathbf{t}}^{2}+\frac{1}{2}\omega_{F_{2}^{+}}^{2}\mathbf{t}^{2}+\frac{1}{2}p_{u}^{2}+\frac{1}{2}\omega_{A_{2}^{-}}^{2}u^{2}\label{eq:-26}\\
 &  & +\frac{1}{2\mathcal{I}}\mathbf{L}^{2}-\frac{\eta_{-}}{2\mathcal{I}}\mathbf{L}\cdot\mathbf{J}_{\mathbf{s}}+\frac{\eta_{-}^{2}}{8\mathcal{I}}\mathbf{J}_{\mathbf{s}}^{2}-\frac{\eta_{+}}{2\mathcal{I}}\mathbf{L}\cdot\mathbf{J}_{\mathbf{t}}+\frac{\eta_{+}^{2}}{8\mathcal{I}}\mathbf{J}_{\mathbf{t}}^{2}.\nonumber 
\end{eqnarray}
where coordinates $\mathbf{s},\mathbf{t},u$ correspond to the vibrations
$F_{2}^{-},F_{2}^{+},A_{2}^{-}$ respectively. As in our analysis
of (\ref{eq:-22}), it will be useful to introduce a total angular
momentum operator $\mathbf{M}=\mathbf{J}_{\mathbf{s}}+\mathbf{J_{t}}-\mathbf{L}$
combining vibrational angular momentum operators $\mathbf{J_{s}},\mathbf{J_{t}}$
with body-fixed angular momentum $\mathbf{L}$. Energy eigenstates
$\Psi$ can be classified by $\mathbf{M}^{2}=\left(\mathbf{J}_{\mathbf{s}}+\mathbf{J_{t}}-\mathbf{L}\right)^{2}$,
$\mathbf{J}_{\mathbf{s}}^{2}$, $\mathbf{J_{t}}^{2}$, $\mathbf{L}^{2}$
and vibrational phonon-numbers $N_{F_{2}^{-}},N_{F_{2}^{+}},N_{A_{2}^{-}}$,
and additionally by their transformation under the $O_{h}$ symmetry
group, where $O_{h}$ acts on a state by transforming the vibrational
coordinates and then performing a compensating rotation of the state:
\begin{equation}
\Psi\rightarrow\rho_{\mathrm{rot}}\left(g\right)\otimes\rho_{\mathrm{vib}}\left(g\right)\Psi.\label{eq:-27}
\end{equation}
Explicitly, this action is generated by
\[
C_{4}:\Psi\rightarrow P_{\mathbf{s}}P_{\mathbf{t}}P_{u}e^{-\frac{2\pi i}{4}\mathbf{n}_{4}\cdot\mathbf{M}}\Psi
\]
\begin{equation}
C_{3}:\Psi\rightarrow e^{-\frac{2\pi i}{3}\mathbf{n}_{3}\cdot\mathbf{M}}\Psi\label{eq:-28}
\end{equation}
\[
-I:\Psi\rightarrow P_{u}P_{\mathbf{s}}\Psi
\]
where $\mathbf{n}_{4}=\begin{pmatrix}0\\
0\\
1
\end{pmatrix}$, $\mathbf{n}_{3}=\frac{1}{\sqrt{3}}\begin{pmatrix}-1\\
-1\\
-1
\end{pmatrix}$ and where $P_{\mathbf{s}},P_{\mathbf{t}},P_{u}$ are parity operators
on the vibrational coordinates. We demand that states transform trivially
under the subgroup $\left\langle C_{4},C_{3}\right\rangle \cong O$
consisting of rotations. Within a fixed $\mathbf{M}^{2}=\left(\mathbf{J}_{\mathbf{s}}+\mathbf{J}_{\mathbf{t}}-\mathbf{L}\right)^{2},\mathbf{J}_{\mathbf{s}}^{2},\mathbf{J}_{\mathbf{t}}^{2},\mathbf{L}^{2},N_{F_{2}^{-}},N_{F_{2}^{+}},N_{A_{2}^{-}}$
sector, we perform character theory calculations and determine $A_{1}$
summands as before. Calculating the resulting spectrum, and then fitting
the frequencies, Coriolis parameters $\eta_{+/-}$ and moment of inertia
$\Lambda$ of the $B=4$ to nuclear data, we obtain the best fit (in
a least-squares sense) for the values
\begin{equation}
\begin{array}{ccc}
\hbar\omega_{F_{2}^{-}}\approx9.7\mathrm{MeV} &  & \frac{\hbar^{2}}{\mathcal{I}}\approx4\mathrm{MeV}\\
\hbar\omega_{F_{2}^{+}}\approx11.7\mathrm{MeV} &  & \eta_{+}\approx0.71\\
\hbar\omega_{A_{2}^{-}}\approx15.1\mathrm{MeV} &  & \eta_{-}\approx0.13
\end{array}.\label{eq:-29}
\end{equation}
In Table \ref{spectrumcube} we list all allowed states up to $30$
MeV for the parameter values in (\ref{eq:-29}). The states in this
energy range consist of both $1$-phonon and $2$-phonon excitations.
With these $6$ parameters we are able to describe $11$ of the $12$
experimentally observed Helium-4 states below $30$ MeV complete with
the correct spin and parity assignments, and we predict one further
$0^{+}$ state at 23.4 MeV. A column $E_{\eta=0}$ is included to
show the spectrum when Coriolis effects are neglected: the Coriolis
corrections have a particularly large effect on the $2^{+}$ states,
raising the energy of the lowest $2^{+}$ excitation by as much as
$3.3$ MeV (comparing favourably to experiment). Note that the ordering
of the fitted frequencies does not agree with that of the Skyrme model
values in Table \ref{vibs}, which put the $E^{+}$ as the lowest-energy
vibration. This discrepancy can perhaps be understood by considering
behaviour beyond small vibrations: recall that the $E^{+}$ vibration
is associated with the breakup of the $B=4$ Skyrmion into two $B=2$
or four $B=1$ Skyrmions. Physically, the breakup energy for $^{4}\text{He}\rightarrow{}^{2}\text{H}+{}^{2}\text{H}$
is $23.8\text{MeV}$, higher than the breakup energy for $^{4}\text{He}\rightarrow{}^{2}\text{H}+\text{p}$
(which should be associated with the $F_{2}$ modes) at $20.3\text{MeV}$.

Our picture suggests that the 20.2 MeV $0^{+}$ state should be identified
with a two-phonon excitation of the $F_{2}^{-}$ mode of the cube.
Promisingly, electron scattering measurements \cite{frosch} of the
transition form factor for the $0^{+}$ suggest collective behaviour,
as noted by the authors of \cite{werntz}. More recent work based
on an ab initio study gives further evidence for the collective interpretation
of this state, suggesting a breathing mode \cite{bacca}. We agree
on the collective nature of this state but, based on the $B=4$ cube,
suggest that the breathing mode should be assigned a higher frequency
than our $F_{2}^{-}$ mode. To compare these two interpretations it
would be worthwhile computing transition form factors from our model.
This would require the explicit form of the Skyrme fields at each
point in our configuration space which, while possible in principle,
is beyond the scope of this paper. There have also been studies of
the negative parity excited states making use of Wigner's theory based
on approximate $SU(4)$ symmetry \cite{walecka}. Our novel picture
has the advantage of giving a unified understanding of almost all
observed excited states, both positive and negative parity, in terms
of simple vibrations of the $B=4$ cube.

\begin{table}[h]
\begin{tabular}{c|c|c|c|c|c|c|c|c|c|c}
$J^{P}$ & $N_{F_{2}^{-}}$ & $N_{F_{2}^{+}}$ & $N_{A_{2}^{-}}$ & $\mathbf{J}_{\mathbf{s}}^{2}$ & $\mathbf{J}_{\mathbf{t}}^{2}$ & $\mathbf{L}^{2}$ & $\mathbf{M}^{2}$ & $E$ & $E_{\mathrm{exp}}$ & $E_{\eta=0}$\tabularnewline
\hline 
$0^{+}$ & 0 & 0 & 0 & 0 & 0 & 0 & 0 & 0 & 0 & 0\tabularnewline
\hline 
$0^{+}$ & 2 & 0 & 0 & 0 & 0 & 0 & 0 & 19.4 & 20.2 & 19.4\tabularnewline
\hline 
$0^{-}$ & 1 & 1 & 0 & 2 & 2 & 0 & 0 & 21.9 & 21.0 & 21.4\tabularnewline
\hline 
$2^{-}$ & 1 & 0 & 0 & 2 & 0 & 6 & 12 & 22.2 & 21.8 & 21.7\tabularnewline
\hline 
$0^{+}$ & 0 & 2 & 0 & 0 & 0 & 0 & 0 & 23.4 & - & 23.4\tabularnewline
\hline 
$1^{-}$ & 1 & 1 & 0 & 2 & 2 & 2 & 0 & 24.2 & 24.3 & 25.4\tabularnewline
\hline 
$2^{+}$ & 0 & 1 & 0 & 0 & 2 & 6 & 12 & 27.0 & 27.4 & 23.7\tabularnewline
\hline 
$1^{+}$ & 1 & 0 & 1 & 2 & 0 & 2 & 0 & 28.3 & 28.3 & 28.8\tabularnewline
\hline 
$1^{-}$ & 0 & 1 & 1 & 0 & 2 & 2 & 0 & 28.5 & 28.4 & 30.8\tabularnewline
\hline 
$2^{-}$ & 1 & 1 & 0 & 2 & 2 & 6 & 0 & 28.9 & 28.4 & 33.4\tabularnewline
\hline 
$0^{-}$ &  &  &  &  &  &  &  &  & 28.6 & \tabularnewline
\hline 
$2^{+}$ & 0 & 2 & 0 & 0 & 6 & 6 & 0 & 28.4 & 28.7 & 35.4\tabularnewline
\hline 
$2^{+}$ & 2 & 0 & 0 & 6 & 0 & 6 & 0 & 29.9 & 29.9 & 31.4\tabularnewline
\hline 
\end{tabular}
\centering{}\caption{Vibrating $B=4$ spectrum up to $30$ MeV.}
\label{spectrumcube}
\end{table}

\section{Vibration-isospin coupling and the $B=7$ Skyrmion}

\begin{figure}[h]
\begin{centering}
\includegraphics[scale=0.3]{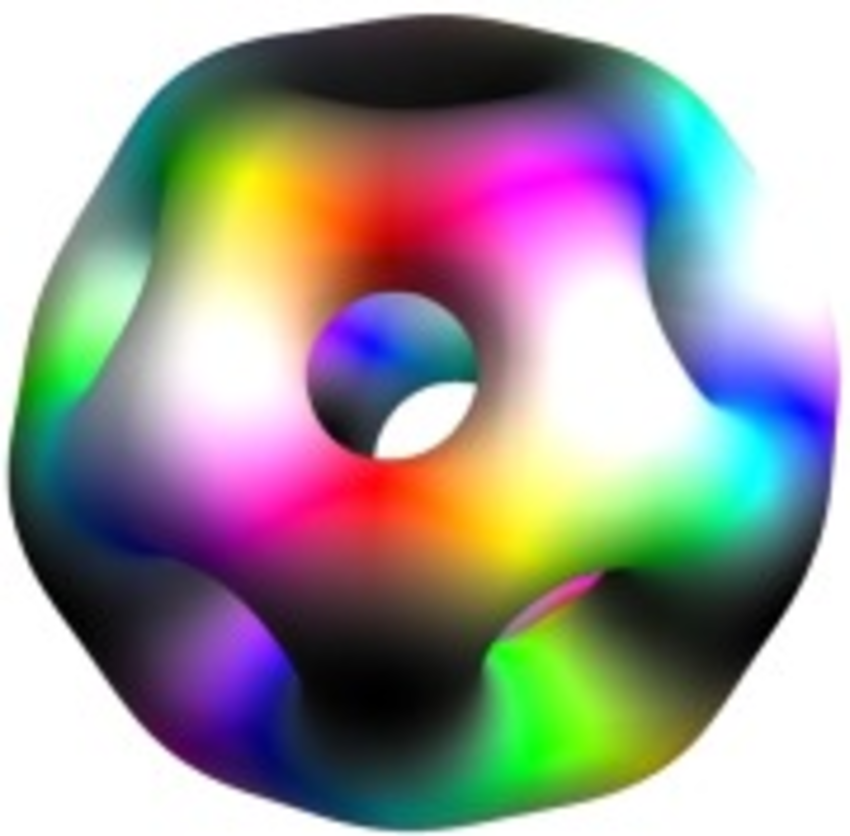}
\par\end{centering}
\caption{B=7 Skyrmion with $I_{h}$ symmetry. Figure courtesy of Chris Halcrow.}

\end{figure}
The lowest-energy $B=7$ Skyrmion is a dodecahedron with symmetry
group $I_{h}$ and its normal modes were studied in detail in \cite{baskerville7}.
If vibrations are not included, the high degree of symmetry of the
$B=7$ means that the lowest energy isospin $\frac{1}{2}$ state has
spin $\frac{7}{2}$. In reality the observed ground state of the Lithium-7/Beryllium-7
isodoublet has spin $\frac{3}{2}$. It was suggested in \cite{halcrow7}
that to capture this state one should include a five-fold degenerate
vibration which transforms in the $H_{g}^{5}$ irrep of $I_{h}$ and
which is generated by pairs of opposite pentagonal faces pulling away
from the center of the Skyrmion. In \cite{halcrowthesis} this vibration
was treated within a harmonic approximation and interactions between
rotations and vibrations were neglected. Here we extend that analysis
to include the Coriolis corrections.

We take our configuration space $\mathcal{C}\simeq SU(2)\times SU(2)\times\mathbb{R}^{5}$
to include the $H_{g}^{5}$ vibration along with rotations and isorotations.
Recalling (\ref{eq:-3}), we should now take $\mathbf{A}_{i}$ to
be a 6-component vector for each $i$ as we are including isorotations.
Introduce coordinates such that the total metric takes the form (to
the order we are interested in)
\begin{equation}
\tilde{g}=\begin{pmatrix}\mathbb{\sigma} & ds_{i}\end{pmatrix}\begin{pmatrix}\Lambda_{0} & \Lambda_{0}\mathbf{A}_{i}\\
\mathbf{A}_{i}^{T}\Lambda_{0} & \delta_{ij}
\end{pmatrix}\begin{pmatrix}\sigma\\
ds_{j}
\end{pmatrix}\label{eq:-17-1}
\end{equation}
with 
\begin{equation}
\Lambda_{0}=\begin{pmatrix}\Lambda_{L}\mathbb{I}_{3} & 0\\
0 & \Lambda_{K}\mathbb{I}_{3}
\end{pmatrix},\label{eq:-18-1}
\end{equation}
$\mathbf{A}_{i}$ linear and vanishing at the equilibrium configuration
$s_{1}=s_{2}=s_{3}=0.$ Recall Jahn's rule from the end of section
3. Now that we are including isorotations, Jahn's rule should be generalised:
the gauge field $\mathbf{A}_{i}$ now corresponds to a singlet of
$I_{h}$ under an action of $I_{h}$ isomorphic to $\rho_{\mathrm{vib}}\otimes\rho_{\mathrm{vib}}\otimes\left(\rho_{\mathrm{rot}}\oplus\rho_{\text{isorot}}\right)$
where $\rho_{\mathrm{isorot}}$ denotes the representation in which
isorotations transform under $I_{h}$. In the present case, rotations
transform as $T_{1g}^{3}$ and isorotations transform as $T_{2g}^{3}$.
An easy calculation shows that 
\begin{equation}
H_{g}^{5}\otimes H_{g}^{5}\otimes\left(T_{1g}^{3}\oplus T_{2g}^{3}\right)\approxeq2A_{g}^{1}\oplus\cdots\label{eq:-30}
\end{equation}
so there is the possibility of non-trivial Coriolis terms coupling
vibrations to spin and isospin (note that we have two copies of the
trivial representation and so the coupling will be determined up to
two arbitrary constants). We wish to find the symmetry-allowed form
of $\mathbf{A}_{i}$, and for this we need explicit coordinates: note
that the usual action of $I_{h}$ on $\mathbb{R}^{3}=\left\langle e_{1},e_{2},e_{3}\right\rangle $
is isomorphic to $T_{1u}^{3}$ and that the symmetric square $T_{1u}^{3}\otimes_{\mathrm{sym}}T_{1u}^{3}\cong A_{g}^{1}\oplus H_{g}^{5}$
contains a copy of the $H_{g}^{5}$ irrep we are interested in. So
we pick vibrational coordinates $s_{1},s_{2},s_{3},s_{4},s_{5}$ (and
conjugate momenta $p_{i}$) such that the action of $I_{h}$ is just
like the action of $I_{h}$ on this $H_{g}^{5}$ subspace with basis
\begin{equation}
\begin{array}{c}
\frac{1}{\sqrt{2}}\left(e_{2}\otimes e_{3}+e_{3}\otimes e_{2}\right),\frac{1}{\sqrt{2}}\left(e_{1}\otimes e_{3}+e_{3}\otimes e_{1}\right),\\
\frac{1}{\sqrt{2}}\left(e_{1}\otimes e_{2}+e_{2}\otimes e_{1}\right),\frac{1}{\sqrt{6}}\left(2e_{3}\otimes e_{3}-e_{1}\otimes e_{1}-e_{2}\otimes e_{2}\right),\\
\frac{1}{\sqrt{2}}\left(e_{1}\otimes e_{1}-e_{2}\otimes e_{2}\right).
\end{array}\label{eq:-33-1}
\end{equation}
In these coordinates we can compute singlets, giving
\begin{eqnarray}
\mathbf{A}_{1} & = & \frac{\eta_{L}}{2\Lambda_{L}}\left(-s_{5}-\sqrt{3}s_{4},-s_{3},s_{2},0,0,0\right)^{T}+\frac{\eta_{K}}{2\Lambda_{K}}\left(0,0,0,-s_{5},-s_{3},2s_{2}\right)^{T}\nonumber \\
\mathbf{A}_{2} & = & \frac{\eta_{L}}{2\Lambda_{L}}\left(s_{3},\sqrt{3}s_{4}-s_{5},-s_{1},0,0,0\right)^{T}+\frac{\eta_{K}}{2\Lambda_{K}}\left(0,0,0,-s_{3},s_{5},-2s_{1}\right)^{T}\nonumber \\
\mathbf{A}_{3} & = & \frac{\eta_{L}}{2\Lambda_{L}}\left(-s_{2},s_{1},2s_{5},0,0,0\right)^{T}+\frac{\eta_{K}}{2\Lambda_{K}}\left(0,0,0,s_{2}-\sqrt{3}s_{4},s_{1},-s_{5}\right)^{T}\label{eq:-31}\\
\mathbf{A}_{4} & = & \frac{\eta_{L}}{2\Lambda_{L}}\left(\sqrt{3}s_{1},-\sqrt{3}s_{2},0,0,0,0\right)^{T}+\frac{\eta_{K}}{2\Lambda_{K}}\left(0,0,0,\sqrt{3}s_{3},\sqrt{3}s_{5},0\right)^{T}\nonumber \\
\mathbf{A}_{5} & = & \frac{\eta_{L}}{2\Lambda_{L}}\left(s_{1},s_{2},-2s_{3},0,0,0\right)^{T}+\frac{\eta_{K}}{2\Lambda_{K}}\left(0,0,0,s_{1},-s_{2}-\sqrt{3}s_{4},s_{3}\right)^{T}\nonumber 
\end{eqnarray}
where $\eta_{L}$ and $\eta_{K}$ are constants. This leads to a Hamiltonian
\begin{eqnarray}
\mathcal{H} & = & \frac{1}{2}\mathbf{p}^{2}+\frac{1}{2}\omega^{2}\mathbf{s}^{2}+\frac{1}{2\Lambda_{L}}\mathbf{L}^{2}+\frac{1}{2\Lambda_{K}}\mathbf{K}^{2}\label{eq:-36}\\
 &  & -\left(\frac{\eta_{L}}{2\Lambda_{L}}\mathbf{L}\cdot\mathbf{J}_{\mathbf{s}}^{L}+\frac{\eta_{K}}{2\Lambda_{K}}\mathbf{K}\cdot\mathbf{J}_{\mathbf{s}}^{K}\right)+\left(\frac{\eta_{L}^{2}}{4\Lambda_{L}}\mathbf{J}_{\mathbf{s}}^{L}\cdot\mathbf{J}_{\mathbf{s}}^{L}+\frac{\eta_{K}^{2}}{4\Lambda_{K}}\mathbf{J}_{\mathbf{s}}^{K}\cdot\mathbf{J}_{\mathbf{s}}^{K}\right).\nonumber 
\end{eqnarray}
involving the vibrational angular momentum operators
\begin{equation}
\mathbf{J}_{\mathbf{s}}^{L}=\begin{pmatrix}-M_{23}+M_{15}+\sqrt{3}M_{14}\\
M_{13}-\sqrt{3}M_{24}+M_{25}\\
-M_{12}-2M_{35}
\end{pmatrix}\label{eq:-34-1}
\end{equation}
and
\begin{equation}
\mathbf{J}_{\mathbf{s}}^{K}=\begin{pmatrix}M_{15}+M_{23}+\sqrt{3}M_{34}\\
M_{13}-M_{25}-\sqrt{3}M_{45}\\
-2M_{12}+M_{35}
\end{pmatrix}\label{eq:-35-1}
\end{equation}
where $M_{ij}=s_{i}p_{j}-s_{j}p_{i}$. $\mathbf{J}_{\mathbf{s}}^{L}$
and $\mathbf{J}_{\mathbf{s}}^{K}$ generate rotations in what is now
a 5-dimensional vibrational space and generalise the vibrational angular
momentum $\mathbf{J}_{\mathbf{s}}$ of (\ref{eq:-21}). We are interested
in eigenstates of (\ref{eq:-36}), which can be classified by $\mathbf{L}^{2}$,
$\mathbf{K}^{2}$ and vibrational phonon-number. Consider one-phonon
states: with respect to a Cartesian basis $\left\{ s_{k}\exp\left(-\alpha\mathbf{s}^{2}\right)\right\} $
of vibrational wavefunctions, it is clear how the $M_{ij}$ act: 
\begin{equation}
M_{ij}s_{k}\exp\left(-\alpha\mathbf{s}^{2}\right)=-i\left(\delta_{il}\delta_{jk}-\delta_{jl}\delta_{ik}\right)s_{l}\exp\left(-\alpha\mathbf{s}^{2}\right)\label{eq:-37}
\end{equation}
and thus how $\mathbf{J}_{\mathbf{s}}^{L}$ and $\mathbf{J}_{\mathbf{s}}^{K}$
act. We diagonalise $\mathcal{H}$ numerically. The relevant group
for imposing the F-R constraints is the universal cover of the icosahedral
group $I$, namely the binary icosahedral group $2I\subset SU\left(2\right)$,
which has presentation
\begin{equation}
\left\langle a,b|\left(ab\right)^{2}=a^{3}=b^{5}\right\rangle .\label{eq:-38}
\end{equation}
F-R constraints tell us that physical states must transform trivially
under the action of the generators $a$ and $b$, given in our coordinates
by
\begin{equation}
s:\Psi\rightarrow e^{\frac{2\pi i}{3}\mathbf{n}_{a}^{L}\cdot\mathbf{L}}\otimes e^{\frac{2\pi i}{3}\mathbf{n}_{a}^{K}\cdot\mathbf{K}}\otimes\rho_{\mathrm{vib}}\left(a\right)\Psi\label{eq:-39}
\end{equation}
\[
t:\Psi\rightarrow e^{\frac{2\pi i}{5}\mathbf{n}_{b}^{L}\cdot\mathbf{L}}\otimes e^{\frac{6\pi i}{5}\mathbf{n}_{b}^{K}\cdot\mathbf{K}}\otimes\rho_{\mathrm{vib}}\left(b\right)\Psi
\]
where 
\[
\mathbf{n}_{a}^{L}=\begin{pmatrix}\sqrt{\frac{2}{15}\left(5-\sqrt{5}\right)}\\
0\\
\sqrt{\frac{1}{15}\left(5+2\sqrt{5}\right)}
\end{pmatrix},\mathbf{n}_{a}^{K}=\begin{pmatrix}-\sqrt{\frac{2}{15}\left(5+\sqrt{5}\right)}\\
0\\
-\sqrt{\frac{1}{15}\left(5-2\sqrt{5}\right)}
\end{pmatrix},\mathbf{n}_{b}^{L}=\begin{pmatrix}0\\
0\\
1
\end{pmatrix},\mathbf{n}_{b}^{K}=\begin{pmatrix}0\\
0\\
1
\end{pmatrix}.
\]
The first few allowed states are listed in Table \ref{spectrumicos-1}
along with the expectation values of the Coriolis terms. These Coriolis
terms represent our corrections to the spectrum found in \cite{halcrowthesis}
which assumed complete separation of rotations and vibrations. That
work focused on the isospin $\frac{1}{2}$ sector: within this sector
one obtains a zero-phonon state with spin $\frac{7}{2}$ (identified
with a $4.6$ MeV excitation of Lithium-7) and one-phonon states with
spins $\frac{3}{2}$, $\frac{5}{2}$, and $\frac{7}{2}$ (identified
with $0$, $6.7$ and $9.7$ MeV excitations of Lithium-7). Ignoring
Coriolis terms, the one-phonon states form a rotational band with
energies following a simple $J\left(J+1\right)$ pattern. The experimental
data doesn't fit this pattern particularly well: the rotational band
energy ratio 
\[
\frac{E\left(J=\frac{7}{2}\right)-E\left(J=\frac{3}{2}\right)}{E\left(J=\frac{5}{2}\right)-E\left(J=\frac{3}{2}\right)}=\frac{7\times9-3\times5}{5\times7-3\times5}=2.4
\]
 which is to be compared with the experimental result $\frac{9.7}{6.7}\approx1.4$.
We now consider the effect of including Coriolis terms for these isospin
$\frac{1}{2}$ states. Recall that, for the $B=7$ Skyrmion, $\frac{\Lambda_{K}}{\Lambda_{L}}\sim0.1$
as found in \cite{manko}. So it is reasonable to assume that, for
the one-phonon states, the most important effect of the Coriolis terms
is the energy splitting of size $\frac{\eta_{K}}{\Lambda_{K}}$ which
(for $\eta_{K}>0$) lowers the energy of the spin $\frac{3}{2}$ state
while raising the energies of the spin $\frac{5}{2}$ and $\frac{7}{2}$
states. We now get 
\[
\frac{E\left(J=\frac{7}{2}\right)-E\left(J=\frac{3}{2}\right)}{E\left(J=\frac{5}{2}\right)-E\left(J=\frac{3}{2}\right)}=\frac{7\times9-3\times5+\frac{8\eta_{K}\Lambda_{L}}{\Lambda_{K}}}{5\times7-3\times5+\frac{8\eta_{K}\Lambda_{L}}{\Lambda_{K}}}
\]
 which reproduces the experimental ratio of $1.4$ for a Coriolis
parameter of $\eta_{K}\approx\frac{25}{4}\frac{\Lambda_{K}}{\Lambda_{L}}\sim0.5$.
It would be interesting to calculate $\eta_{K}$ explicitly from the
Skyrme model and compare to this value.

We have learnt from this example that, in situations where the isospin
moment of inertia is much smaller than the spin moment of inertia,
it is perfectly possible for the isospin Coriolis corrections to compete
with the usual $\frac{1}{2\Lambda_{L}}J\left(J+1\right)$ rotational
band splittings. This kind of effect is particularly important for
odd $B$ Skyrmions like the $B=7$, where non-zero isospin is inevitable
(isospin taking half-integer values). This fits with the fact that
the rotational band picture has been much more successful for even
$B$ nuclei than for odd $B$ nuclei.

\begin{table}[h]
\begin{centering}
\begin{tabular}{c|c|c}
Spin/Isospin & Energy without Coriolis terms & Coriolis terms\tabularnewline
\hline 
$\left(\frac{7}{2}_{\mathrm{}}/\frac{1}{2}\right)_{\mathrm{0-phonon}}$ & $\frac{5}{2}\hbar\omega+\frac{\hbar^{2}}{2\Lambda_{L}}\frac{7}{2}\frac{9}{2}+\frac{\hbar^{2}}{2\Lambda_{K}}\frac{1}{2}\frac{3}{2}$ & $0$\tabularnewline
\hline 
$\left(\frac{3}{2}/\frac{1}{2}\right)_{\mathrm{1-phonon}}$ & $\frac{7}{2}\hbar\omega+\frac{\hbar^{2}}{2\Lambda_{L}}\frac{3}{2}\frac{5}{2}+\frac{\hbar^{2}}{2\Lambda_{K}}\frac{1}{2}\frac{3}{2}$ & $3\frac{\eta_{L}}{2\Lambda_{L}}-\frac{3}{2}\frac{\eta_{K}}{2\Lambda_{K}}+6\frac{\eta_{L}^{2}}{4\Lambda_{L}}+6\frac{\eta_{K}^{2}}{4\Lambda_{K}}$\tabularnewline
\hline 
$\left(\frac{5}{2}/\frac{1}{2}\right)_{\mathrm{1-phonon}}$ & $\frac{7}{2}\hbar\omega+\frac{\hbar^{2}}{2\Lambda_{L}}\frac{5}{2}\frac{7}{2}+\frac{\hbar^{2}}{2\Lambda_{K}}\frac{1}{2}\frac{3}{2}$ & $\frac{1}{2}\frac{\eta_{L}}{2\Lambda_{L}}+\frac{\eta_{K}}{2\Lambda_{K}}+6\frac{\eta_{L}^{2}}{4\Lambda_{L}}+6\frac{\eta_{K}^{2}}{4\Lambda_{K}}$\tabularnewline
\hline 
$\left(\frac{7}{2}/\frac{1}{2}\right)_{\mathrm{1-phonon}}$ & $\frac{7}{2}\hbar\omega+\frac{\hbar^{2}}{2\Lambda_{L}}\frac{7}{2}\frac{9}{2}+\frac{\hbar^{2}}{2\Lambda_{K}}\frac{1}{2}\frac{3}{2}$ & $-3\frac{\eta_{L}}{2\Lambda_{L}}+\frac{\eta_{K}}{2\Lambda_{K}}+6\frac{\eta_{L}^{2}}{4\Lambda_{L}}+6\frac{\eta_{K}^{2}}{4\Lambda_{K}}$\tabularnewline
\hline 
$\left(\frac{3}{2}_{\mathrm{}}/\frac{3}{2}\right)_{\mathrm{\mathrm{0}-phonon}}$ & $\frac{5}{2}\hbar\omega+\frac{\hbar^{2}}{2\Lambda_{L}}\frac{3}{2}\frac{5}{2}+\frac{\hbar^{2}}{2\Lambda_{K}}\frac{3}{2}\frac{5}{2}$ & $0$\tabularnewline
\hline 
$\left(\frac{1}{2}/\frac{3}{2}\right)_{\mathrm{1-phonon}}$ & $\frac{7}{2}\hbar\omega+\frac{\hbar^{2}}{2\Lambda_{L}}\frac{1}{2}\frac{3}{2}+\frac{\hbar^{2}}{2\Lambda_{K}}\frac{3}{2}\frac{5}{2}$ & $-\frac{3}{2}\frac{\eta_{L}}{2\Lambda_{L}}+3\frac{\eta_{K}}{2\Lambda_{K}}+6\frac{\eta_{L}^{2}}{4\Lambda_{L}}+6\frac{\eta_{K}^{2}}{4\Lambda_{K}}$\tabularnewline
\hline 
$\left(\frac{3}{2}/\frac{3}{2}\right)_{\mathrm{1-phonon}}$ & $\frac{7}{2}\hbar\omega+\frac{\hbar^{2}}{2\Lambda_{L}}\frac{3}{2}\frac{5}{2}+\frac{\hbar^{2}}{2\Lambda_{K}}\frac{3}{2}\frac{5}{2}$ & $-3\frac{\eta_{L}}{2\Lambda_{L}}-3\frac{\eta_{K}}{2\Lambda_{K}}+6\frac{\eta_{L}^{2}}{4\Lambda_{L}}+6\frac{\eta_{K}^{2}}{4\Lambda_{K}}$\tabularnewline
\hline 
\end{tabular}
\par\end{centering}
\caption{Spectrum including one-phonon $H_{g}^{5}$ excitations of the $B=7$
Skyrmion.}
\label{spectrumicos-1}
\end{table}

\section{Conclusions and further work}

We have developed a model of Helium-4 based on $F_{2}$ and $A_{2}$
vibrations of the cubic $B=4$ Skyrmion. Our model includes interactions
between rotations and vibrations in the form of Coriolis terms. The
spectrum gives a good match to the experimental data, with the Coriolis
terms significantly improving the fit. The lowest state not captured
by the model is a $0^{-}$ state at $28.6$ MeV, and we predict one
so far unobserved $0^{+}$ state at $23.4$ MeV. We have also extended
these ideas to the $B=7$ Skyrmion, clarifying the role of isospin-vibration
coupling.

The example in section 5 suggests a general feature which should occur
in vibrational quantization of Skyrmions with non-zero isospin. It
has been noted (e.g. in \cite{su4}) that for large $B$ the isospin
moments of inertia for Skyrmions are much smaller than the spin moments
of inertia, with $\Lambda_{K}\sim B$ and $\Lambda_{L}\sim B^{\frac{5}{3}}$.
So, for large $B$, isospin Coriolis corrections can become more important
than the usual $\frac{1}{2\Lambda_{L}}J\left(J+1\right)$ rotational
band splittings. In fact the stable large nuclei all have large isospin
and so these ideas are important for the Skyrme model description
of many real nuclei.

It would be interesting to calculate the actual values for the Coriolis
coefficients $\eta$ numerically within the Skyrme model. This requires
explicit field configurations for vibrating Skyrmions but such configurations
have been calculated before in e.g. \cite{halcrow}. It would also
be interesting to study the effect of the gauge field $\mathbf{A}_{i}$
for a situation in which shape space includes large deformations (not
just small vibrations).

Finally it should be noted that, while our ideas have been outlined
within the context of the Skyrme model, this work is very general
and these ideas could be applied to other soliton systems in which
one is interested in the interplay between zero and non-zero modes.

\section*{Acknowledgements}

I am grateful to my supervisor Professor Nick Manton for many useful
discussions and guidance, and to Chris Halcrow for suggesting I apply
these techniques to the $B=7$ Skyrmion. I am supported by an EPSRC
studentship. This work has been partially supported by STFC consolidated
grant ST/P000681/1.

\end{document}